\documentstyle[aps,preprint]{revtex}

\widetext
\draft
\tighten
\begin{document}

\preprint{EFUAZ FT-95-14}

\title{Interactions of a $j=1$ Boson in the $2(2j+1)$
Component Theory\thanks{Submitted to ``{\it International  Journal
of Theoretical Physics}".}}

\author{{\bf Valeri V. Dvoeglazov}\thanks{On leave of absence from
{\it Dept. Theor. \& Nucl. Phys., Saratov State University,
Astrakhanskaya ul., 83, Saratov\, RUSSIA.}\,
Email: dvoeglazov@main1.jinr.dubna.su}}

\address {
Escuela de F\'{\i}sica, Universidad Aut\'onoma de Zacatecas \\
Antonio Doval\'{\i} Jaime\, s/n, Zacatecas 98000, ZAC., M\'exico\\
Internet address:  VALERI@CANTERA.REDUAZ.MX
}

\date{April, 1995}

\maketitle

\begin{abstract}
The amplitudes for boson-boson and fermion-boson interactions are
calculated in the second order of perturbation theory in
the Lo\-ba\-chev\-sky space.
An essential ingredient of the used model is the Weinberg's
$2(2j+1)$ component formalism for describing a particle
of spin $j$, recently developed substantially.
The boson-boson amplitude is then compared
with the two-fermion amplitude obtained long ago by Skachkov
on the ground of the hamiltonian formulation of quantum field
theory on the mass hyperboloid, $p_0^2 -{\bf p}^2=M^2$, proposed
by Kadyshevsky. The pa\-ra\-met\-ri\-za\-tion of the amplitudes by means
of the momentum transfer in the Lo\-ba\-chev\-sky space leads to
same spin structures in the expressions of $T$ matrices
for the fermion and the boson cases. However,  certain
differences are found. Possible physical applications are discussed.
\end{abstract}


\newpage

The scattering amplitude for the two-fermion interaction had been
obtained in the Lobachevsky space in the second order of perturbation
theory  long ago~[1a,Eq.(31)]:
\begin{eqnarray}\label{eq:TF}
\lefteqn{T^{(2)}_V ({\bf k} (-) {\bf p}, {\bf p}) =
-g_v^2 \frac{4m^2}{\mu^2 +4 \text{{\bf \ae}}^{\,2}} -
4g_v^2\frac{(\bbox{\sigma}_1 \text{{\bf \ae}})(\bbox{\sigma}_2
\text{{\bf \ae}}) - (\bbox{\sigma}_1 \bbox{\sigma}_2)
\text{{\bf \ae}}^2}{\mu^2 +4\text{{\bf \ae}}^{\,2}} -\nonumber}\\
&-& {8g_v^2 p_0 \text{\ae}_0 \over m^2}\,
\frac{i\bbox{\sigma}_1 [{\bf p} \times \text{{\bf \ae}} ] +i\bbox{\sigma}_2
[{\bf p} \times \text{{\bf \ae}} ]}{\mu^2 +4 \text{{\bf \ae}}^{\,2}} -
{8g_v^2 \over m^2}\,\frac{p_0^2 \text{\ae}_0^2 +2p_0 \text{\ae}_0 ({\bf p}
\cdot \text{{\bf \ae}}) - m^4}{\mu^2 +4\text{{\bf \ae}}^{\,2}} -\nonumber\\
&-& \frac{8g_v^2}{m^2}\,\frac{(\bbox{\sigma}_1 {\bf p})
(\bbox{\sigma}_1 \text{{\bf \ae}}) (\bbox{\sigma}_2 {\bf p})
(\bbox{\sigma}_2 \text{{\bf \ae}})}{\mu^2 +4\text{{\bf \ae}}^{\,2}}\quad,
\end{eqnarray}
$g_v$ is the coupling constant.
This treatment  is based on the use of the formalism
of separation of the Wigner rotations and parametrization
of currents by means of the Pauli-Lyuban'sky vector, developed in
the sixties~\cite{Chesh}. The quantities
$$\text{\ae}_0 = \sqrt{\frac{m(\Delta_0 +m)}{2}}\quad,\quad
\text{{\bf \ae}} = {\bf n}_\Delta \sqrt{\frac{m(\Delta_0 -m)}{2}}$$
are the components of  the 4-vector of a momentum ``half-transfer".
This concept is closely connected with a notion of the half-velocity
of a particle~\cite{Chernik}.
The 4-vector $\Delta_{\mu}$:
\begin{mathletters}
\begin{eqnarray}
\bbox{\Delta} &=& \bbox{\Lambda}^{-1}_{{\bf p}} {\bf k}
= {\bf k} (-) {\bf p} = {\bf k}
-\frac{{\bf p}}{m} (k_0 - \frac{{\bf k}\cdot
{\bf p}}{p_0 +m})\quad,\\
\Delta_0 &=& (\Lambda^{-1}_{p} k)_0 = (k_0 p_0
-{\bf k}\cdot{\bf p})/m = \sqrt{m^2\,+ \,\bbox{\Delta}^2}
\end{eqnarray}
\end{mathletters}
could be regarded as the momentum transfer vector in
the Lobachevsky space.\footnote{I keep  a notation and  a terminology
of ref.~\cite{Skachkov}. {\it E.g.}, the vector current with
taking into account the Pauli term is
\begin{equation}
j^\mu_{\sigma\sigma^\prime} ({\bf p}, {\bf k}) =
\overline u_\sigma ({\bf p})\left \{ g_v \gamma^\mu - f_v
\frac{\sigma^{\mu\nu}}{2m} q_\nu \right \} u_{\sigma^\prime} ({\bf
k})\quad, \quad  q=p-k
\end{equation}
and
\begin{equation}\label{chalf}
j^\mu_{\sigma\sigma^\prime} ({\bf p}, {\bf k}) =
\sum_{\sigma_p=-1/2}^{1/2} j^\mu_{\sigma\sigma_p} ({\bf k} (-) {\bf p},
{\bf p}) \,\,D^{\slantfrac{1}{2}}_{\sigma_p \sigma^\prime}
\left \{ V^{-1} (\Lambda_p,k)\right \}\quad,
\end{equation}
where
$D^{\slantfrac{1}{2}}_{\sigma_p
\sigma^\prime} \left \{ V^{-1} (\Lambda_p, k)\right \} =\xi_{\sigma_p}
D^{\slantfrac{1}{2}}\left \{ V^{-1} (\Lambda_p,k)\right
\}\xi_{\sigma^\prime}$;\, $D^{\slantfrac{1}{2}} (A) \equiv
D^{(\slantfrac{1}{2},\,0)} (A)$ is the Wigner matrix of the irreducible
representation of the $SU (2)$ group (or rotation group).  The technique
of construction of $D^J (A)$  could be found in
ref.~\cite[p.51,70,English edition]{Novozh}.}$^{,}\,$
\footnote{In general,
for each particle in interaction
one should understand under 4-momenta $p^\mu_i$ and $k^\mu_i$\, $(i=1,2)$
their covariant generalizations, $\overcirc{p}^\mu_i$, $\overcirc{k}^\mu_i$,
{\it e.g.}, refs.~\cite{Chesh,Faustov,Dvoegl2}:
$$\overcirc{{\bf k}} = (\bbox{\Lambda}_{{\cal P}}^{-1} {\bf k}) =
{\bf k} - \frac{\bbox{{\cal P}}}{\sqrt{{\cal P}^2}} \left ( k_0 -
\frac{\bbox{{\cal P}}\cdot{\bf k}}{ {\cal P}_0 + \sqrt{{\cal P}^2}}
\right )\quad,$$
$$\overcirc{k}_0 = (\Lambda_{{\cal P}}^{-1} k)_0 =
\sqrt{m^2 +\overcirc{{\bf k}}^{\,\,2}},$$
with ${\cal P}= p_1 +p_2$,
$\Lambda_{{\cal P}}^{-1} {\cal P} = ({\cal M},\, \bbox{0})$.
However, we omit the circles above the momenta
in the following, because  in the case under consideration
we do not miss physical information
if use the corresponding quantities in c.m.s.,
${\bf p}_1 = -{\bf p}_2 = {\bf p}$ and ${\bf k}_1 =-{\bf k}_2 = {\bf k}$.}

This amplitude had been  used for physical applications in
the framework of the Kadyshevsky's version of the quasipotential
approach~\cite{Kadysh,Skachkov}.

{}From the other hand, in  ref.~\cite{Joos,Weinberg}
an attractive $2(2j+1)$ component formalism for describing
particles of higher spins has been proposed.
As opposed to the Proca 4-vector potentials, that transform according
to the $({1\over 2}, {1\over 2})$ representation of
the Lorentz group, the spinor $2(2j+1)$ component functions are constructed
via the representation $(j,0)\oplus (0,j)$ in the Joos-Weinberg formalism.
This way of description of higher spin particles is on an equal
footing to the description of the Dirac spinor particle,
whose wave function transforms according to the
$({1\over 2},0)\oplus (0, {1\over 2})$ representation.
The $2(2j+1)$- component analogues of the Dirac functions in
the momentum space are\footnote{These functions obey
the orthonormalization equations,
${\cal U}^\dagger ({\bf p})\gamma_{00}\,{\cal U} ({\bf p})= M $,
$M$ is the mass of
the Joos-Weinberg particle. The similar normalization condition
exists for  $ {\cal V} ({\bf p})$, the functions of
``negative-energy states".}
\begin{equation}\label{pos}
{\cal U} ({\bf p})= \sqrt{{M\over 2}} \left (\matrix{
D^J \left (\alpha({\bf p})\right )\xi_\sigma\cr
D^J \left (\alpha^{-1\,\dagger}({\bf p})\right )\xi_\sigma\cr
}\right )\quad,
\end{equation}
for the positive-energy states; and
\begin{equation}\label{neg}
{\cal V} ({\bf p})= \sqrt{{M\over 2}} \left (\matrix{
D^J \left (\alpha({\bf p})\Theta_{[1/2]}\right )\xi^*_\sigma\cr
D^J \left ( \alpha^{-1\,\dagger}({\bf p}) \Theta_{[1/2]}\right
)(-1)^{2J}\xi^*_\sigma\cr }\right )\quad,
\end{equation} for the
negative-energy states, ref.~\cite[p.107]{Novozh}, with the following
notations being used:
\begin{equation}
\alpha({\bf p})=\frac{p_0+M+(\bbox{\sigma}
\cdot{\bf p})}{\sqrt{2M(p_0+M)}},\quad
\Theta_{[1/2]}=-i\sigma_2\quad.
\end{equation}
For instance, in the case of spin $j=1$, one has
\begin{mathletters}
\begin{eqnarray}
&&D^{\,1}\left (\alpha({\bf p})\right ) \,=\,
1+\frac{({\bf J}\cdot{\bf p})}{M}+
\frac{({\bf J}\cdot{\bf p})^2}{M(p_0+M)}\quad,\\
&&D^{\,1}\left (\alpha^{-1\,\dagger}({\bf p})\right ) \,=\,
1-\frac{({\bf J}\cdot{\bf p})}{M}+
\frac{({\bf J}\cdot{\bf p})^2}{M(p_0+M)}\quad,  \\
&&D^{\,1}\left (\alpha({\bf p}) \Theta_{[1/2]}\right ) \,=\,
\left [1+\frac{({\bf J}\cdot{\bf p})}{M}+
\frac{({\bf J}\cdot{\bf p})^2}{M(p_0+M)}\right ]\Theta_{[1]}\quad, \\
&&D^{\,1}\left (\alpha^{-1\,\dagger}({\bf p}) \Theta_{[1/2]}\right ) \,=\,
\left [1-\frac{({\bf J}\cdot{\bf p})}{M}+
\frac{({\bf J}\cdot{\bf p})^2}{M(p_0+M)}\right ]\Theta_{[1]}\quad,
\end{eqnarray}
\end{mathletters}
($\Theta_{[1/2]}$,\,$\Theta_{[1]}$ are the Wigner operators for spin 1/2
and 1, respectively).
In spite of some antiquity of this formalism, in our opinion, it does not
deserve to be retired. From the phenomenological viewpoint
this approach provides the necessary framework of constructing
a QCD-based effective field theory of higher-spin hadronic resonances
and could yield new insights into the quark structure of these
excited hadrons. Recently, much attention has been paid to this
formalism~\cite{DVA0,DVA00} (see also the papers~\cite{Sankar,DVO0}
regarding similar problems). Unfortunately,
all the authors of the old classic works
devoted to this formalism have missed the possibility of another
definition of negative-energy bispinors ${\cal V} ({\bf p}) =
S^c_{[1]} \,{\cal U} ({\bf p}) \,\equiv \,{\cal C}_{[1]} \,
{\cal K}\, {\cal U} ({\bf p}) \,\sim \,\gamma_5 {\cal U} ({\bf p})$,
like the Dirac $j=1/2$ case.\footnote{We don't
treat here the new Majorana-like constructs in the
$(j,0)\oplus (0,j)$ representation space~\cite{DVAM} referring
the reader to our recent works~\cite{DVOM}.} $S^c_{[1]}$ is the
charge conjugation operator for $j=1$, ref.~[10c]; ${\cal K}$ is the
operation of complex conjugation.  This definition, based on the use of
another form of the  Ryder-Burgard relation~[10c,d],
leads to different physical content: in the latter case a boson and its
antiboson have opposite relative intrinsic parities (like Dirac spinor
particles). This is an example of another class of Poincar\`e
invariant theories (the Bargmann-Wightman-Wigner-type quantum field
theories~\cite{Wigner}).  This remarkable fact, which has been proven in
refs.~[10c,d], hints that  the problem of the adequate choice
of the field operator has profound  physical significance.

In refs.~\cite{Weinberg,Hammer,Novozh,Dvoegl,Dvoegl1}
the Feynman diagram technique was
discussed in the above-mentioned six-component
formalism for particles of spin $j=1$. The following
Lagrangian:\,\footnote{In the following I prefer to use
the Euclidean metric because this metric got application in
a lot of papers on the $2(2j+1)$ formalism.}$^,$\footnote{The
expression for the Lagrangian has been generalized in refs.~\cite{DVO0}.
In this paper we are still going to use the previous one in order to
emphasize features of the formalism relevant to the purposes of
the work.}
\begin{eqnarray}
\lefteqn{{\cal L}=\overline{\Psi}(x)\Gamma_{\mu\nu}\loarrow{\nabla}_{\mu}
\roarrow{\nabla}_{\nu}\Psi(x) - M^2\overline{\Psi}(x)\Psi(x)-{1 \over
4}F_{\mu\nu}F_{\mu\nu}+}\nonumber\\
&+&\frac{e\lambda}{12}F_{\mu\nu}\overline{\Psi}(x)\gamma_{5,\mu\nu}
\Psi(x)+\frac{e\kappa}{12 M^2}\partial_{\alpha}F_{\mu\nu}\overline{\Psi}(x)
\gamma_{6,\mu\nu,\alpha\beta}\nabla_{\beta}\Psi(x)
\end{eqnarray}
has been  used there.
In the above formula we have
$\tensor\nabla_{\mu}=-i\partial_{\mu}\mp eA_{\mu}$;
$F_{\mu\nu}=\partial_{\mu}A_{\nu}-\partial_{\nu}A_{\mu}$ is the
 electromagnetic field tensor; $A_{\mu}$ is the 4-vector of
electromagnetic field; $\overline{\Psi}, \Psi$  are
the six-component wave functions
(WF) of a massive $j=1$ Joos-Weinberg particle.
The following expression has been obtained
for the interaction vertex of the particle with a
photon described by the vector potential, ref.~\cite{Hammer,Dvoegl}:
\begin{equation}
-e\Gamma_{\alpha\beta}(p+k)_{\beta} - {ie\lambda \over
6}\gamma_{5,\alpha\beta}q_{\beta}+{e\kappa \over
6M^2}\gamma_{6,\alpha\beta,\mu\nu}q_{\beta}q_{\mu}(p+k)_{\nu}\quad,
\label{22}
\end{equation}
where $\Gamma_{\alpha\beta}=\gamma_{\alpha\beta}+\delta_{\alpha\beta}$;\,
$\gamma_{\alpha\beta}$; \,$\gamma_{5, \alpha\beta}$; \,
$\gamma_{6,\alpha\beta, \mu\nu}$ \, are the $6\otimes 6$-matrices which
have been  described in ref.~\cite{Barut,Weinberg}:
\begin{mathletters}
\begin{eqnarray}
\gamma_{ij}\,&=&\,\pmatrix{
0 & \delta_{ij}\openone - J_i J_j- J_j J_i \cr
\delta_{ij}\openone - J_i J_j- J_j J_i & 0 \cr
}\quad,\\
\gamma_{i4}\,&=&\,\gamma_{4i}=\pmatrix{
0 & iJ_i \cr
-iJ_i & 0 \cr
}\quad,\quad
\gamma_{44}=\pmatrix{
0 & \openone \cr
\openone & 0\cr
}\quad,
\end{eqnarray}
\end{mathletters}
and
\begin{mathletters}
\begin{eqnarray}
\gamma_{5,\alpha\beta}&=&i [\gamma_{\alpha\mu}, \gamma_{\beta\mu}]_-\quad,\\
\gamma_{6,\alpha\beta,\mu\nu}&=&
[\gamma_{\alpha\mu},\gamma_{\beta\nu}]_{+}
+2\delta_{\alpha\mu}\delta_{\beta\nu}-[\gamma_{\beta\mu},
\gamma_{\alpha\nu}]_{+}-2\delta_{\beta\mu}\delta_{\alpha\nu}\quad.
\end{eqnarray}
\end{mathletters}
$J_i$ are the  spin matrices for a $j=1$  particle,
$e$ is  the electron charge, $\lambda$ and $\kappa$ are the quantities that
correspond to the magnetic dipole moment and the electric quadrupole moment,
respectively.

In order to obtain the 4-vector current for the interaction
of a Joos-Weinberg boson with the external field
one can use the known formulas of refs.~\cite{Skachkov,Chesh}, which
are valid for any spin:
\begin{equation}
{\cal U}^\sigma({\bf p}) =
\bbox{S}_{{\bf p}} \,{\cal U}^\sigma({\bf 0})\quad, \quad \bbox{S}_{{\bf
p}}^{-1} \bbox{S}_{{\bf k}} = \bbox{S}_{{\bf k}(-){\bf p}}\cdot I\otimes
D^{1}\left \{ V^{-1}(\Lambda_p, k)\right \}\quad,
\end{equation}
\begin{equation}
W_\mu({\bf p})\cdot D\left \{ V^{-1}(\Lambda_{p}, k)\right \}
= D\left \{ V^{-1}(\Lambda_{p}, k)\right \}
\cdot\left [ W_\mu({\bf k})
+\frac{p_\mu+k_\mu}{M(\Delta_0+M)}p_\nu W_\nu ({\bf k})\right ],
\end{equation}
\begin{equation}
k_\mu W_\mu ({\bf p})\cdot D\left \{ V^{-1}(\Lambda_{p}, k)\right \} =
-D\left \{ V^{-1}(\Lambda_{p}, k)\right \}\cdot p_\mu W_\mu ({\bf k})
\quad.
\end{equation}
$W_\mu$ is the Pauli-Lyuban'sky 4-vector of relativistic spin.
The matrix $D^J \left \{ V^{-1} (\Lambda_{p}, k)\right \}$
is written for spin $j=1$ as follows:
\begin{eqnarray}
\lefteqn{D^{(j=1)}\left \{ V^{-1}(\Lambda_{p}, k)\right \}=
\frac{1}{2M(p_0+M)(k_0+M)(\Delta_0+M)} \left \{
\left [{\bf p}\times {\bf k}\right ]^2+\right.}\nonumber\\
&+&\left.\left [(p_0+M)(k_0+M) -{\bf k}\cdot{\bf p}\right ]^2 +
2i\left [(p_0+M)(k_0+M)-{\bf k}\cdot{\bf p}\right ] \left \{
{\bf J}\cdot\left [{\bf p}\times {\bf k}\right ]\right \}-\right.\nonumber\\
&-&\left.2\{{\bf J}\cdot\left
[{\bf p}\times{\bf k}\right ]\}^2\right \}\quad.
\end{eqnarray}
However, the formulas obtained in
ref.~\cite{Dvoegl1}:\footnote{Attention is drawn to the definition
of $\gamma_5$ matrix which differs by a sign from the definition
used in refs.~[17c,19].}
\begin{mathletters}
\begin{eqnarray}
\bbox{S}_{{\bf p}}^{-1} \gamma_{\mu\nu} \bbox{S}_{{\bf p}}\,
&=& \,\gamma_{44} \left \{ \delta_{\mu\nu} - {1\over M^2}
\chi_{[\mu\nu]} ({\bf p})\otimes \gamma_5 - {2\over M^2}
\Sigma_{[\mu\nu]} ({\bf p})\right \}\quad,\\
\bbox{S}_{{\bf p}}^{-1} \gamma_{5,\mu\nu} \bbox{S}_{{\bf p}}\,
&=&\, 6i \left \{ - {1\over M^2}
\chi_{(\mu\nu)} ({\bf p})\otimes \gamma_5 + {2\over M^2}
\Sigma_{(\mu\nu)} ({\bf p})\right \}\quad,
\end{eqnarray}
\end{mathletters}
where
\begin{mathletters}
\begin{eqnarray}
\chi_{[\mu\nu]} ({\bf p}) \,&=&\, p_\mu W_\nu ({\bf p})
+ p_\nu W_\mu ({\bf p})\quad,\\
\chi_{(\mu\nu)} ({\bf p}) \, &=&\, p_\mu W_\nu ({\bf p})
- p_\nu W_\mu ({\bf p})\quad, \\
\Sigma_{[\mu\nu]} ({\bf p}) \,&=&\,{1\over 2}
\left \{ W_\mu ({\bf p}) W_\nu ({\bf p}) +
W_\nu ({\bf p}) W_\mu ({\bf p})\right \}\quad,\\
\Sigma_{(\mu\nu)} ({\bf p}) \,&=&\,{1\over 2}
\left \{ W_\mu ({\bf p}) W_\nu ({\bf p}) -
W_\nu ({\bf p}) W_\mu ({\bf p}) \right \}\quad,
\end{eqnarray}
\end{mathletters}
lead to the 4- current of a $j=1$ Joos-Weinberg
particle more directly:\footnote{{\it Cf.} with a $j=1/2$
case:
\begin{mathletters}
\begin{eqnarray}
&&\bbox{S}_p^{-1} \gamma_\mu \bbox{S}_p = {1 \over m}
\gamma_0 \left \{ \openone \otimes p_\mu + 2\gamma_5 \otimes W_\mu ({\bf
p}) \right \} \quad, \\
&&\bbox{S}_p^{-1} \sigma_{\mu\nu} \bbox{S}_p =
- {4\over m^2} \openone \otimes \Sigma_{(\mu\nu)} ({\bf p})
+ {2\over m^2} \gamma_5 \otimes \chi_{(\mu\nu)}
({\bf p})\quad,
\end{eqnarray}
\end{mathletters}
and, then,
\begin{equation}
j_\mu^{\sigma_p\nu_p} ({\bf k} (-) {\bf p}, {\bf p}) =
{1\over m} \xi^\dagger_{\sigma_p} \left \{2g_v \text{\ae}_0 p_\mu
+ f_v \text{\ae}_0 q_\mu + 4 g_{{\cal M}}  W_\mu ({\bf p})
(\bbox{\sigma} \cdot \text{{\bf \ae}})\right \} \xi_{\nu_p}\quad,
\quad (g_{{\cal M}} = g_v + f_v)\quad.\label{current}
\end{equation}
The indices $p$ indicate that  the Wigner rotations have been separated
out and, thus,  all spin indices have been ``resetted"
on the momentum ${\bf p}$. One can re-write~[1b] the
electromagnetic current~(\ref{current}):
\begin{equation}
j_\mu^{\sigma_p\nu_p} ({\bf k}, {\bf p}) =
- {e\,m\over \text{\ae}_0} \xi^\dagger_{\sigma_p} \left \{
g_{{\cal E}} (q^2)\, (p+k)^\mu +
g_{{\cal M}} (q^2)\,
\left [{1\over m} W_\mu ({\bf p}) (\bbox{\sigma}\cdot \bbox{\Delta})
- {1\over m} (\bbox{\sigma}\cdot \bbox{\Delta}) W_\mu ({\bf p})\right ]
\right \} \xi_{\nu_p}\quad.\label{current1}
\end{equation}
$g_{{\cal E}}$ and $g_{{\cal M}}$ are the analogues of the Sachs electric
and magnetic form factors.
Thus, if we regard $g_{S,T,V}$ as  effective coupling
constants depending on the momentum transfer one can ensure ourselves
that the form of the currents for a spinor particle  and
for a $j=1$ boson is the same (with Wigner rotations separated out).}
\begin{mathletters}
\begin{eqnarray}
j_{\mu}^{\sigma_{p}\nu_{p}}({\bf p}, {\bf k}) &=&
j_{\mu \,(S)}^{\sigma_{p}\nu_{p}}({\bf p}, {\bf k}) +
j_{\mu \,(V)}^{\sigma_{p}\nu_{p}}({\bf p}, {\bf k}) +
j_{\mu \,(T)}^{\sigma_{p}\nu_{p}}({\bf p}, {\bf k})\quad,\\
j_{\mu \,(S)}^{\sigma_{p}\nu_{p}}({\bf p}, {\bf k}) \,&=&\,
-\,g_S \xi^\dagger_{\sigma_p} \left \{  (p+k)_\mu \left (
1+ \frac{({\bf J}\cdot \bbox{\Delta})^2}{M (\Delta_0 + M)} \right )\right
\} \xi_{\nu_p}\quad,\label{curs}\\
j_{\mu \,(V)}^{\sigma_{p}\nu_{p}}({\bf
p}, {\bf k}) \,&=&\, -\,g_V \xi^\dagger_{\sigma_p} \left \{ (p+k)_{\mu}+
{1\over M}W_{\mu}({\bf p})({\bf J}\cdot\bbox{\Delta})- {1\over M}({\bf
J}\cdot\bbox{\Delta}) W_{\mu}({\bf p})\right \} \xi_{\nu_p}\quad,
\label{cur}\\
j_{\mu \,(T)}^{\sigma_{p}\nu_{p}}({\bf p}, {\bf k}) \,&=&\,
-\, g_T \xi_{\sigma_p}^\dagger \left \{ - (p+k)_\mu
\frac{({\bf J}\cdot \bbox{\Delta})^2}{M (\Delta_0 + M)}+
\right.\label{curt}\\
&& \left. \qquad\qquad\qquad + {1\over M} W_{\mu}({\bf p})({\bf
J}\cdot\bbox{\Delta})- {1\over M}({\bf J}\cdot\bbox{\Delta}) W_{\mu}({\bf
p})\right \} \xi_{\nu_p}\quad.\nonumber
\end{eqnarray}
\end{mathletters}
Let us note an interesting feature~[10b,13c]. The 6-spinors
${\cal U} ({\bf p})$ and ${\cal V} ({\bf p})$ defined by Eqs.
(\ref{pos},\ref{neg}) do not form a complete set:
\begin{equation}
{1\over M} \left \{{\cal U} ({\bf p}) \overline {\cal U} ({\bf p})
+{\cal V} ({\bf p}) \overline {\cal V} ({\bf p}) \right \} \,=\,
\pmatrix{\openone & \bbox{S}_{{\bf p}}\otimes \bbox{S}_{{\bf p}} \cr
\bbox{S}^{-1}_{{\bf p}}\otimes \bbox{S}^{-1}_{{\bf p}} &
\openone\cr}\quad.
\end{equation}
But, if  regard ${\cal V}_2 ({\bf p})=
\gamma_5 {\cal V} ({\bf p})$ one can obtain
the complete set. Fortunately,
\begin{equation}
\overline {\cal V}_2 ({\bf 0}) {\cal U}_1 ({\bf 0}) = 0\quad,
\end{equation}
what  permits us to keep the parametrization (\ref{chalf}).
As a matter of fact,
in Eqs. (\ref{curs}-\ref{curt}) we have used the second definition
of negative energy spinors, ref.~[10c,d].

Next, let me  now represent the Feynman matrix
element corresponding to the diagram of  two-boson interaction, mediated
by the particle described by the vector
potential,  in the form~\cite{Skachkov,Dvoegl} (read the remark in
the footnote \# 2):
\begin{eqnarray}
\lefteqn{ <p_1, p_2;\, \sigma_1, \sigma_2\vert \hat T^{(2)}
\vert k_1, k_2;\, \nu_1, \nu_2>
=}\nonumber\\
&=&\sum^{1}_{\sigma_{ip}, \nu_{ip}, \nu_{ik} =-1} D^{\dagger\quad
(j=1)}_{\sigma_1\sigma_{1p}} \left \{V^{- 1} (\Lambda_{\cal P}, p_1)\right
\} D^{\dagger\quad (j=1)}_{\sigma_2\sigma_{2p}} \left \{V^{-1}
(\Lambda_{\cal P}, p_2)\right \}\times\nonumber\\
&\times&T^{\nu_{1p}\nu_{2p}}_{\sigma_{1p}\sigma_{ 2p}}({\bf k} (-)
{\bf p}, {\bf  p}) D^{(j=1)}_{\nu_{1p}\nu_{1k}}\left \{V^{-1}
(\Lambda_{p_1}, k_1)\right \} D^{(j=1)}_{\nu_{1k}\nu_1}\left \{V^{-1}
(\Lambda_{\cal P}, k_1)\right \}\times\nonumber\\
&\times& D^{(j=1)}_{\nu_{2p}\nu_{2k}} \left\{ V^{-1} (\Lambda_{p_2},
k_2)\right \} D^{(j=1)}_{\nu_{2k}\nu_2}\left\{ V^{-1} (\Lambda_{\cal P},
k_2)\right \}\quad,
\end{eqnarray}
where
\begin{equation} \label{ampl}
T^{\nu_{1p}\nu_{2p}}_{\sigma_{1p}\sigma_{2p}} ({\bf k}(-) {\bf p}, {\bf p}) =
\xi^\dagger_{\sigma_{1p}} \xi^\dagger_{\sigma_{2p}} \,
T^{(2)} ({\bf k} (-) {\bf p},\, {\bf p})\, \xi_{\nu_{1p}} \xi_{\nu_{2p}}\quad,
\end{equation}
$\xi^\dagger$, $\xi$ are the analogues of  Pauli spinors.
The calculation of the amplitude (\ref{ampl}) yields
($p_0 = -ip_4$,\,\,$\Delta_0 = -i \Delta_4$):
\begin{eqnarray}\label{212}
\lefteqn{\hat T^{(2)} ({\bf k}(-){\bf p}, {\bf p})
\,=\, g^2 \left\{ \frac{\left [p_0
(\Delta_0 +M) + ({\bf p}\cdot \bbox{\Delta})\right ]^2
-M^3 (\Delta_0+M)}{M^3 (\Delta_0 -M)}+\right.}\nonumber\\
&+&\left.\frac{i ({\bf J}_1+{\bf J}_2)\cdot\left [{\bf p}
\times\bbox{\Delta}\right ]}
{\Delta_0-M}\left [ \frac{p_0 (\Delta_0 +M)+{\bf p}\cdot
\bbox{\Delta}}{M^3} \right ]
+ \frac{({\bf J}_1\cdot \bbox{\Delta})({\bf
J}_2\cdot \bbox{\Delta})-({\bf J}_1\cdot{\bf J}_2) \bbox{\Delta}^2}{2M
(\Delta_0-M)}-\right.\nonumber\\
&-&\left.\frac{1}{M^3}\frac{{\bf J}_1\cdot\left [{\bf p}
\times\bbox{\Delta}\right ] \,\,{\bf J}_2\cdot \left [{\bf
p} \times\bbox{\Delta}\right ]}{\Delta_0-M}\right\}\quad.
\end{eqnarray}
We have assumed $g_S = g_V = g_T$ above, what is motivated
by group-theoretical reasons and by the analogy discussed in
the footnote \# 8. The expression (\ref{212}) reveals  the
advantages of the $2(2j+1)$- formalism, since
it looks like  the amplitude for the interaction of two spinor particles
with the substitutions
$$\frac{1}{2M(\Delta_0 - M)}
\Rightarrow\frac{1}{\bbox{\Delta}^2}\quad \mbox{and}
\quad {\bf J}\Rightarrow \bbox{\sigma}\quad.$$
The calculations hint that many analytical results produced for
a Dirac fermion could be applicable to describing a $2(2j+1)$
particle. Nevertheless, it is required adequate explanation
of the obtained difference. An inquisitive reader could note:
its origin lies at the kinematical
level.  Free-space (without interaction) Joos-Weinberg equations admit
acausal tachyonic solutions~\cite{DVA00}. ``Interaction
introduced in the massive Weinberg equations will couple to both the
causal and acausal solutions and thus cannot give physically
reasonable results". However, let us not forget that we have used
the Tucker-Hammer approach~[17b], indeed, that does not possess
tachyonic solutions.\footnote{I am not going to deal further
with this subject in the present paper.
The description of dynamics based on new
kinematical ground~\cite{DVA0} will be given in a detailed
publication.}

For the sake of completeness I also present the amplitudes for
interaction of $j=1$ and $j=0$ particles, $j=1/2$ and $j=0$ particles, and
$j=1/2$ and $j=1$ particles.  Let us use the equation for the 4-current of
spinor particle ($f_v=0$), defined by the formula (\ref{current1}); the
equation (\ref{curs}) with ${\bf J}=0$, for a scalar particle ({\it e.g.},
ref.~\cite{Rohrlich}); the
equation (\ref{cur}) for the 4- current of a $j=1$ particle in the
Joos-Weinberg  formalism.  Following to the technique of
"resetting"  polarization indices, we obtain in a first case:
\begin{eqnarray}
\lefteqn{\hat T^{(2)}_{0\,1} ({\bf k}, {\bf p}) \,=\,
-\,\frac{g_0 g_1}{2m_1(\Delta_1^0 - m_1)}
\left \{ - 2m_1^2 (\Delta_1^0 +m_1)+ \right.}\\
&+&\left .\left [ p_1^0 (\Delta_1^0 +m_1) +
({\bf p}\cdot {\bbox\Delta}_1)\right ] (p_1^0+ p_2^0 + k_1^0 + k_2^0)
+ i{\bf J}\cdot
[{\bf p}\times {\bbox\Delta}_1]\, (p_1^0 +p_2^0 +k_1^0 +k_2^0)\right
\}\quad,\nonumber
\end{eqnarray}
what has a similar form to
$\hat T^{(2)}_{0\slantfrac{1}{2}} ({\bf k}, {\bf p})$, which is below.

As a result of lengthy calculations one can write the boson-fermion
amplitudes in the following form:
\begin{eqnarray}
\lefteqn{
\hat T^{(2)}_{0\slantfrac{1}{2}} ({\bf k}, {\bf p}) \,=\,
-\,\frac{2g_0 g_{\slantfrac{1}{2}}}{(2m_1)^{3/2}
(\Delta_1^0 - m_1) \sqrt{\Delta_1^0 +m_1}}
\left \{ - 2m_1^2 (\Delta_1^0 +m_1)+ \right.}\\
&+&\left . \left [ p_1^0 (\Delta_1^0 +m_1) +
({\bf p}\cdot {\bbox\Delta}_1)\right ] (p_1^0+ p_2^0 + k_1^0 + k_2^0)
+ i\bbox{\sigma}\cdot
[{\bf p}\times \bbox{\Delta}_1] (p_1^0 +p_2^0 +k_1^0 +k_2^0)\right
\}\quad.\nonumber
\end{eqnarray}

\medskip

and

\medskip

\begin{eqnarray}
\lefteqn{\hat
T^{(2)}_{1\,\slantfrac{1}{2}} ({\bf p}, {\bf k}) \,=\, -\,
\frac{2g_1 g_{\slantfrac{1}{2}}}{(2m_1)^{3/2}
(\Delta_1^0 - m_1) \sqrt{\Delta_1^0 +m_1}}
\left \{ - 2m_1^2 (\Delta_1^0 +m_1)+ \right.}\nonumber\\
&+&\left. \left [ p_{1}^{0} (\Delta_1^0 +m_1) + ({\bf p}\cdot{\bbox\Delta}_1)
\right ] \left ( p_{1}^{0} + p_{2}^{0} + k_{1}^{0} + k_{2}^{0}\right )
+ i \bbox{\sigma}_1 \cdot [{\bf p}\times \bbox{\Delta}_1]
(p_{1}^{0}+p_{2}^{0}+k_{1}^{0}+k_{2}^{0}) -\right.\nonumber\\
&-&\left. {i\over m_2}{\bf J}_2\cdot [{\bf p} \times \bbox{\Delta}_2]
\left [(p_{1}^{0}+p_{2}^{0}) (\Delta_1^0 + m_1) +
\frac{({\bf p}\cdot\bbox{\Delta}_1) (p_{1}^{0}+p_{2}^{0}+m_1+m_2)^2}{2
(p_{1}^{0}+m_1)(p_{2}^{0}+m_2)}\right ] - \right.\nonumber\\
&-& \left. m_1 \left [(\bbox{\sigma}_1\cdot \bbox{\Delta}_2)({\bf J}_2\cdot
\bbox{\Delta}_1) -(\bbox{\sigma}_1\cdot{\bf J}_2)(\bbox{\Delta}_1\cdot
\bbox{\Delta}_2) + i {\bf J}_2 \cdot[\bbox{\Delta}_1\times
\bbox{\Delta}_2]\right ] +\right.\\
&+&\left. \bbox{\sigma}_1 \cdot[{\bf p}\times
\bbox{\Delta}_1]\,\,{\bf J}_2\cdot [{\bf p}\times
\bbox{\Delta}_2] \frac{(p_{1}^{0}+p_{2}^{0}+m_1+m_2)^2}{2m_2
(p_{1}^{0}+m_1)(p_{2}^{0}+m_2)}\right \}\quad.\nonumber
\end{eqnarray}
Above we have used the notation:
\begin{mathletters}
\begin{eqnarray}
\bbox{\Delta}_1 &=& {\bf k}-\frac{{\bf p}}{m_1}\left ( k_{1}^{0}-
\frac{{\bf k}\cdot{\bf p}}{p_{1}^{0}+m_1}\right )\quad,\quad
\Delta_{1}^{0}=\sqrt{\bbox{\Delta}^2_1+m_1^2}\quad,\\
\bbox{\Delta}_2 &=& {\bf k}-\frac{{\bf p}}{m_2}
\left ( k_{2}^{0}-\frac{{\bf k}\cdot{\bf p}}{p_{2}^{0}+m_2}\right
)\quad,\quad
\Delta_{2}^{0}=\sqrt{\bbox{\Delta}^2_2+m_2^2} \quad,
\end{eqnarray}
\end{mathletters}
and
$p_{1}^{0}=\sqrt{{\bf p}\,^2+m_1^2}$,
$k_{1}^{0}=\sqrt{{\bf k}\,^2+m_1^2}$, $p_{2}^{0}=\sqrt{{\bf
p}\,^2+m_2^2}$, $k_{1}^{0}=\sqrt{{\bf k}\,^2+m_2^2}$\,.

\bigskip

\bigskip

{\it Discussion and possible physical applications}:

\smallskip

The main result of this paper is the boson-boson amplitude
calculated in the framework of the Joos-Weinberg theory.
The separation of the Wigner rotations permits us to reveal
certain similarities with a $j=1/2$ case. Thus, this result
provide a ground for the conclusion: if
existence of the Joos-Weinberg bosons would be
proven\footnote{As already mentioned in this paper and in refs.~[10c,d],
probably, the crucial experiment for the Joos-Weinberg
boson  could be  based on the
determination of relative intrinsic parities of a boson
and its antiboson.} many calculations produced earlier for fermion-fermion
interactions mediated by the vector potential could be applicable
to processes involving this matter structure. Moreover,
the main result of the paper gives a certain hope at a possibility
of the unified description of fermions and bosons.
Now, I realize that all the above-mentioned is
not surprising.  The principal features of describing the particle world
on the base of relativistic quantum field theory are not in some
special representation of the group, $(1/2,0)\oplus (0,1/2)$
or $(1,0)\oplus (0,1)$, or $(1/2,1/2)$, but in the Lorentz group itself.
Several old papers, {\it e.g.}, ref.~\cite{Ohmura},
and recent paper~\cite{Bruce} can support this conclusion.
However, the difference between denominators
of the amplitudes necessitates us to undertake further study of the
$(1/2,0)\oplus (0,1/2)$ and $(1,0)\oplus (0,1)$ representations.
These representations, of course, are contained in the general
scheme of Joos and Weinberg.

After an appearance of the
paper~\cite{DVA00,DVO0} (see also ref.~\cite{Evans}) we seem to be forced
to use equations of this approach:\footnote{See ref.~[10c,d]
for a discussion on the possible additional term $\wp_{u,v}$
at the mass term for integer spins.}
\begin{equation}
\left [ \gamma^{\mu_1 \mu_2 \ldots
\mu_{2j}}
\partial_{\mu_1}\partial_{\mu_2}\ldots\partial_{\mu_{2j}}+M^{2j}\right
]\Psi(x)=0
\end{equation}
for describing higher-spin particles. In the framework
of the standard  Proca/Rarita-Schwinger approach
we deal with many contradictions in the particle
interpretation of the field transforming on the
corresponding representations of the Lorentz group
({\it e.g.}, acausal solutions even at the kinematical level;
the absence of the well-defined massless limit; the non-consistency
after introduction of interactions; the longitudity
of antisymmetric tensor field after quantization,
what contradicts with the classical limit and with the Weinberg theorem;
longitudinal non-propagating solutions of the Maxwell's equations which
could lead to speculations on a ``massive" photon, ``tired"
light~\cite{Evans} and the
lost of renormalizability of modern quantum field models {\it etc}).

Secondly, to the moment it was realized that a
gluon can be described as a massive particle
with  dynamical mass appearing due to existence
of  the color charge and  the self-interaction.
This treatment permits one to eliminate some
contradictions in the results
of calculations of the proton form factor and the
effective coupling constant $\alpha_S(q^2)$
on the basis of QCD (for recent discussion see ref.~\cite{Field}).
Therefore, the presented amplitude could serve as a base for
describing the gluonium, the bound state of two massive gluons.
The fermion-boson amplitudes could be applied to describing
the quark-diquark  composite system.

For thirty years the quasipotential approach to quantum
field theory~\cite{LT,Kadysh} are regarded to be the most convenient and
sufficiently general formalism for calculation of energy spectra of
composite states.  In the Bethe-Salpeter approach one has a non-physical
parameter (relative time), difficulties with the normalization of the
bound state wave function {\it etc}. All this necessitates
us to introduce
constraints on the wave function. As a matter of fact, they lead to the
approaches which are  equivalent to the quasipotential one.
For recent discussion
see ref.~\cite{Crater,DVOKA}. Therefore, one can use equations
for the equal-time (quasipotential) wave function to achieve
the goals discussed above. {\it E.g.}, for
a composite system formed by fermion and boson of non-equal mass
the equation was given (in the Kadyshevsky version of quasipotential
approach) in~\cite{Link}:
\begin{eqnarray}
\lefteqn{2 \overcirc{p}_2^0 \left ({\cal M} - \overcirc{p}_1^0
-\overcirc{p}_2^0 \right ) \Phi_{\sigma_1
\sigma_2}(\overcirc{{\bf p}}) = }\nonumber\\
\,&=&\, \frac{1}{(2\pi)^3}
\sum_{\nu_1 \nu_2} \int \frac{d^{3} \overcirc{{\bf k}}}{2\overcirc{k}_1^0}
\,\,V^{\nu_1 \nu_2}_{\sigma_1 \sigma_2} (\overcirc{{\bf p}},
\overcirc{{\bf k}}) \,\Phi_{\nu_1 \nu_2}
(\overcirc{{\bf k}})\quad.
\end{eqnarray}
Several works dealing with  phenomenological description
of hadrons in the $(j,0)\oplus (0,j)$ framework have
been published~\cite{DVO-old2,DVO-old3} and submitted
for publication~\cite{DVO-pr}.

Finally,  not having any intentions to shadow theories based
on the concept of vector potential, in our opinion,
the principal question is not yet solved.
It is not in formal advantages of
one or another formalism for describing $j=1$ (or higher spin)
particles, but in ``the nature of Nature's mesons".

\acknowledgments
I  appreciate very much discussions with Prof.
D. V.  Ahluwalia, Prof. A. F. Pashkov and Prof. Yu. F. Smirnov.
I should thank  Prof. N. B. Skachkov for his help in analyzing
several topics.

I am grateful to Zacatecas University for professorship.


\begin{references}

\footnotesize{
\bibitem{Skachkov} N. B. Skachkov,  TMF  22 (1975) 213
[English translation: Theor. Math. Phys.,p.149]; ibid  25 (1975) 313
[Theor. Math. Phys.,p.1154];  N. B. Skachkov and I. L. Solovtsov,
Fiz. Elem. Chast. At. Yadr.  9   (1978)  5 [Sov. J. Part. and Nucl.,p.1]

\bibitem{Chesh} Yu. M. Shirokov, ZhETF 21  (1951) 748;  DAN
SSSR  99 (1954) 737, in Russian;  ZhETF  33  (1957) 1196, 1208 [Sov.
Phys.  JETP  6 (1958) 919, 929]; ZhETF  35   (1958) 1005 [Sov. Phys. JETP
8  (1959) 703]; Chou Kuang Chao and
M. I. Shirokov , ZhETF  34  (1958) 1230 [Sov. Phys. JETP  7  (1958) 851];
A. A. Cheshkov  and  Yu. M. Shirokov,  ZhETF 42  (1962) 144  [Sov. Phys.
JETP  15  (1962) 103];  ZhETF  44  (1963) 1982 [Sov. Phys. JETP  17
(1963) 1333];  A. A. Cheshkov,  ZhETF  50  (1966)  144  [Sov. Phys. JETP
23  (1966)  97]; V. P. Kozhevnikov, V. E. Troitski\u{\i},
S. V. Trubnikov and Yu. M. Shirokov, Teor. Mat. Fiz. 10 (1972) 47

\bibitem{Chernik} N. A. Chernikov, ZhETF 33 (1957) 541 [Sov. Phys. JETP
6 (1958) 422]; Fiz. Elem. Chast. At. Yadra 4 (1973) 773

\bibitem{Novozh} Yu. V. Novozhilov, {\it Vvedenie v teoriyu
elementarnykh chastitz.} (Moscow,  Nauka, 1971)
[{\it Introduction to
Elementary Particle Theory.} (Pergamon Press, Oxford, 1975)]

\bibitem{Faustov} R. N. Faustov, Ann. Phys. (USA)  78
(1973)  176

\bibitem{Dvoegl2} V. V. Dvoeglazov {\it et al.},
Yadern. Fiz. 54 (1991) 658 [Sov. J. Nucl. Phys. , p.398]

\bibitem{Kadysh} V. G. Kadyshevsky,  Nuovo
Cim.  55A (1968) 233;  Nucl. Phys. B6 (1968) 125; V. G. Kadyshevsky, R.
M. Mir-Kasimov  and N. B. Skachkov, Fiz. Elem. Chast. At. Yadra 2 (1972)
635  [Sov. J.  Part. Nucl.,p.69]

\bibitem{Joos} H. Joos,  Fortschr. Phys. 10 (1962) 65

\bibitem{Weinberg}  S.  Weinberg,  Phys. Rev. B133 (1964) 1318;
ibid 134 (1964) 882;  ibid  181 (1969) 1893

\bibitem{DVA0}  D. V. Ahluwalia and D. J. Ernst,
Phys. Lett. B287  (1992) 18;  Int. J.
Mod. Phys. E2 (1993) 397; D.~V.~Ahluwalia, M. B. Johnson and T. Goldman,
Phys. Lett. B316 (1993) 102; D. V. Ahluwalia and T. Goldman,
Mod. Phys. Lett. A8 (1993) 2623

\bibitem{DVA00}  D. V. Ahluwalia and D. J. Ernst,
Mod. Phys. lett. A7 (1992) 1967

\bibitem{Sankar} A. Sankaranarayanan and R. H. Good, jr.,
Nuovo Cim.  36 (1965) 1303;  Phys. Rev. 140B (1965) 509;
A. Sankaranarayanan, Nuovo Cim. 38 (1965) 889

\bibitem{DVO0} V. V. Dvoeglazov, {\it Mapping Between Antisymmetric
Tensor  and Weinberg Formulations.} Preprint EFUAZ  FT-94-05
(hep-th/9408077), Aug.  1994; {\it  What   Particles Are  Described by
the Weinberg  Theory?} Preprint EFUAZ FT-94-06 (hep-th/9408146), Aug.
1994; {\it The Weinberg Propagators.}
Preprint EFUAZ FT-94-07 (hep-th/9408176), Aug. 1994; {\it Can
the $2(2S+1)$ Component Weinberg-Tucker-Hammer Equations Describe
the Electromagnetic Field?} Preprint EFUAZ FT-94-09 (hep-th/9410174),
Oct. 1994, submitted to ``J. Phys. A"

\bibitem{DVAM} D. V. Ahluwalia, M. B. Johnson and T. Goldman,
{\it Space-Time Symmetries: P and CP Violation.}
To be published in ``Proceedings of
the III Int. Wigner Symp. Oxford, 1993"; Mod. Phys. Lett.  A9
(1994) 439; Acta Phys. Polon. B25 (1994) 1267

\bibitem{DVOM} V. V. Dvoeglazov, {\it A Note on the Majorana Theory
for $j=1/2$ and $j=1$ Particle States.} Preprint EFUAZ
FT-94-10 (hep-th/9504157).  Reported at the XVIII Oaxtepec Symp. on Nucl.
Phys.  M\'exico, 1995; {\it Neutral Particles in Light of the
Majorana-Ahluwalia Ideas.} Preprint EFUAZ FT-95-11 (hep-th/9504158),  Feb.
1995

\bibitem{Wigner} E. P. Wigner, in {\it Group Theoretical Concepts and
Methods in Elementary Particle Physics.} Ed. F. G\"ursey.
(Gordon and Breach, 1962)

\bibitem{Hammer}  C. L. Hammer, S. C. McDonald and D. L. Pursey,
Phys. Rev. 171  (1968) 1349; R. H. Tucker and C. L.
Hammer,  Phys. Rev. D 3 (1971)  2448; see also
D. Shay and R. H. Good,  Phys. Rev. 179  (1969)  1410

\bibitem{Dvoegl} V. V. Dvoeglazov and N. B. Skachkov, JINR Communications
P2-84-199 (1984); ibid  P2-87-882 (1987),  in Russian

\bibitem{Dvoegl1}  V. V. Dvoeglazov  and  N. B.  Skachkov, Yadern. Fiz. 48
 (1988) 1770 [Sov. J. of Nucl.  Phys.,p.1065]

\bibitem{Barut} A. O. Barut, I. Muzinich and D. Williams, Phys. Rev.
130 (1963) 442

\bibitem{Rohrlich} F. Rohrlich, Phys. Rev. 80 (1950) 666

\bibitem{Ohmura} T. Ohmura, Prog. Theor. Phys. 16 (1956) 684, 685

\bibitem{Bruce} S. Bruce, Nuovo Cim. 110B (1995) 115

\bibitem{Evans} M. W. Evans, Found. Phys. 24 (1994) 1519, 1671

\bibitem{Field} J. H. Field, Int. J. Mod. Phys. A9 (1994);
M. Consoli and J. H. Field, Phys. Rev. D49 (1994) 1293

\bibitem{LT} A. A. Logunov and A. N. Tavkhelidze,
Nuovo Cim. 29 (1963) 380

\bibitem{Crater} H. W. Crater {\it et al.}, Phys. Rev. D46 (1992) 5117

\bibitem{DVOKA} V. V. Dvoeglazov, Yu. N. Tyukhtyaev and R. N. Faustov,
Mod. Phys. Lett. A8 (1993) 3263; Fiz. Elem. Chast. At. Yadra 25 (1994) 144
[Phys. Part. Nucl.,p.58]

\bibitem{Link} A. D. Linkevich, V. I. Savrin and N. B. Skachkov,
Yadern. Fiz. 37 (1983) 391 [Sov. J. Nucl. Phys.,p.235]

\bibitem{DVO-old2} V. V.  Dvoeglazov  and S. V. Khudyakov, Izvestiya
VUZov:fiz. No. 9  (1994) 110 [Russian  Phys. J. 37 (1994)]

\bibitem{DVO-old3} V. V. Dvoeglazov, Rev. Mex. Fis. Suppl. (Proc. XVII
Oaxtepec Symp.  on Nucl.  Phys., Jan. 4-7, 1994, M\'exico) 40 (1994) 352

\bibitem{DVO-pr} V. V. Dvoeglazov, S. V. Khudyakov and S. B. Solganik,
{\it Relativistic Covariant Equal-Time Equation for Quark-Diquark System.}
Preprint IFUNAM FT-93-24 (hep-ph/9308305), M\'exico, Aug. 1993; V. V.
Dvoeglazov and S. V. Khudyakov, {\it Gluonium as Bound State of Massive
Gluons Described by the Joos-Weinberg Wave Functions.} Preprint IFUNAM
FT-93-35 (hep-ph/9311347), M\'exico, Nov. 1993
}
\end{references}
\end{document}